# Substrate Sensitivity of Monolayer WS$_2$

*Kathleen M. McCreary,[1]* Aubrey T. Hanbicki,[1] Simranjeet Singh,[2] Roland K. Kawakami,[2] Berend T. Jonker[1]*

[1] *Naval Research Laboratory,* Washington DC 20375, USA

[2] *Department of Physics, The Ohio State University*, Columbus OH 43210, USA

* Author Information:

Correspondence and requests for materials should be addressed to K.M.M. (email: kathleen.mccreary@nrl.navy.mil)

We report on preparation dependent properties observed in monolayer WS$_2$ samples synthesized via chemical vapor deposition (CVD) on a variety of common substrates (Si/SiO$_2$, sapphire, fused silica) as well as samples that were transferred from the growth substrate onto a new substrate. The as-grown CVD materials (as-WS$_2$) exhibit distinctly different optical properties than transferred WS$_2$ (x-WS$_2$). In the case of CVD growth on Si/SiO$_2$, following transfer to fresh Si/SiO$_2$ there is a ~50 meV shift of the ground state exciton to higher emission energy in both photoluminescence emission and optical reflection. This shift is indicative of a reduction in tensile strain by ~0.25%. Additionally, the excitonic state in x-WS$_2$ is easily modulated between neutral and charged exciton by exposure to moderate laser power, while such optical control is absent in as-WS$_2$ for all growth substrates investigated. Finally, we observe dramatically different laser power-dependent behavior for as-grown and transferred WS$_2$. These results demonstrate a strong sensitivity to sample preparation that is important for both a fundamental understanding of these novel materials as well as reliable reproduction of device properties.

The novel properties of graphene have stimulated research in other two-dimensional (2D) materials such as hexagonal boron nitride and the transition metal dichalcogenides (TMDs). The TMDs have a chemical composition of $MX_2$ (where M=Mo, W and X=S, Se), and monolayer building blocks of these materials are composed of three sheets of atoms where a top and bottom chalcogen layer encapsulate the center metal sheet. While much of the early work focused on monolayer $MoS_2$, the closely related $WS_2$ exhibits superior optical properties[1–3] and larger spin-orbit coupling,[4] motivating further extensive investigation. $WS_2$ exhibits a strong sensitivity to layer number, with the band structure transitioning from an indirect-gap semiconductor for bulk $WS_2$ to direct-gap semiconductor at monolayer thickness.[1] Fundamental investigations and prototype devices incorporating atomically thin $WS_2$ have demonstrated reasonable electronic mobility,[5,6] high optical responsivity,[1] robust device performance following repeated bending,[7] high valley polarization,[8–10] and long spin lifetimes.[11,12] These properties make $WS_2$ a promising material for a variety of applications including photodetection, flexible electronics, and spintronics.

Monolayers of TMDs can be obtained using a variety of methods including mechanical and chemical exfoliation, atomic layer deposition (ALD),[13] molecular beam epitaxy (MBE),[14] and chemical vapor deposition (CVD).[15] Other innovative options continue to be explored. The CVD approach has emerged as a particularly reliable route to obtain uniform, high-quality, large-area samples,[2,16,17] with wafer-scale synthesis of $WS_2$ recently demonstrated.[18] Due to the elevated temperatures necessary for CVD synthesis, rigid and thermally robust substrates such as $SiO_2$ or $Al_2O_3$ are commonly used.

The ability to transfer films off the growth substrate provides a means to incorporate TMDs with a more diverse array of materials, and is highly desirable for both fundamental investigations and applications. For instance, transfer to high-k dielectrics may provide a means to modify excitonic properties such as binding energies, position, and radii, as Coulomb interactions are sensitive to the local environment.[19,20] The construction of van der Waals heterostructures (vdwh) through sequential stacking of 2D materials is also expected to open the door to new device functionality, and is a rapidly progressing area of research [21] with recent reports of interlayer excitons,[22] ultrafast charge separation,[23] and long valley lifetimes[24] in TMD based vdwh. While reliable transfer techniques will certainly be important as the field continues to grow, there are limited reports that investigate if and how $WS_2$ is impacted.[25-27] Additionally, such studies typically transfer between dissimilar materials (i.e. sapphire to $Si/SiO_2$), making it difficult to determine the source of any observed modifications.    Therefore, it is important to first understand and evaluate any changes brought about by transfer processes between identical substrates so that future investigations may accurately isolate, identify and assess modifications resulting from interactions with various materials. In this letter we directly compare fundamental properties of CVD synthesized monolayer $WS_2$ between the as-grown state (as-$WS_2$) and after removal from the growth substrate and transfer to a fresh supporting substrate (x-$WS_2$).

**Results and Discussion**

Monolayer tungsten disulfide is synthesized at a growth temperature of 825 °C in a 2-inch diameter quartz tube furnace. Powdered $WO_3$ (Alfa Aesar 13398) and sulfur (Alfa Aesar 10755) serve as precursors for the synthesis. Additional details can be found in reference 16. Typically $Si/SiO_2$ (275nm) is employed as growth substrate and is the main focus of this manuscript. However, monolayer $WS_2$ has also been realized on c-plane sapphire and fused silica substrates, and will be briefly discussed. Growth on $Si/SiO_2$ results in multiple triangular islands exhibiting lateral dimensions ranging from a few to several tens of microns, as shown in the optical image, Fig. 1(a). The equilateral triangle morphology is characteristic of single-crystal growth.[28] Examination of a $WS_2$ triangle using AFM (Fig. 1(b)) shows uniform thickness across the triangle as well as a step height of ~0.8 nm (inset of Fig. 1(b)), consistent with monolayer $WS_2$.

Raman and photoluminescence (PL) spectroscopy measurements are performed using a commercial Horiba LabRam confocal spectrometer in ambient atmosphere at room temperature using 532 nm laser excitation unless otherwise stated. A 100× objective focuses the laser beam to a spot with ~1 μm diameter. Raman spectroscopy is a quick and reliable method for layer identification in $WS_2$.[29,30] In addition to the in-plane ($E^1_{2g}$) and out-of-plane ($A^1_g$) first order optical modes, 532 nm laser excitation results in second-order resonant peaks involving longitudinal acoustic modes (LA(M)). A typical Raman spectrum from as-$WS_2$ is show in Fig. 1(c). A multi-lorentzian fit clearly resolves the individual components, with $E^1_{2g}$ and $A^1_g$ peaks located at 355.7 cm$^{-1}$ and 418.4 cm$^{-1}$, respectively. A PL spectrum (Fig. 1(d)), also measured on as-WS2, shows a single peak having maximum intensity at ~1.96 eV, confirming the single layer nature. While sample-to-sample variation can influence the exact PL peak position, repeated measurements obtained across multiple samples as well as separate monolayer regions on the same sample typically exhibit peak emission intensity within ~10 meV of 1.96 eV for monolayer WS$_2$ synthesized on Si/SiO$_2$(275nm).

Two distinct methods are used to remove WS$_2$ monolayers from their growth substrates and transfer them to a fresh Si/SiO$_2$ substrate. The first method is referred to as "PMMA transfer" and involves a wet etching process based on previous techniques used for transferring CVD graphene grown on copper.[31] This process is illustrated in Fig. 2(a) and is described in the Methods. Optical images of PMMA x-WS$_2$ (Fig. 2(a), right panel) demonstrate the shape and uniformity of the flakes are maintained after transfer. The second transfer method is free of wet etchants and ensures a clean substrate-sample interface. This method, termed "PDMS stamp", is illustrated in Fig. 2(b) and is also described in the Methods. This technique provides microscopic precision for both placement and rotation angle of the transferred WS$_2$. The optical image in Fig. 2(b) (right panel) displays a typical PDMS transferred region. The majority of the sample exhibits uniform contrast, although at the edges there are small patches where WS$_2$ is absent.

PL, reflectivity and Raman spectra obtained from representative as-WS$_2$ and PMMA x-WS$_2$ are presented in Fig. 3. The spectrum for as-WS$_2$ (Fig. 3(a)) exhibits a peak PL emission at 1.956 eV, with a FWHM of 35 meV. A considerable blue shift of

~50 meV is evident in the x-WS$_2$ spectrum, where the emission peak shifts to 2.011 eV and the FWHM is reduced to 27 meV. Control samples were fabricated by coating as-WS$_2$ with PMMA then removing it. These samples show no shift in peak position, ruling out doping from PMMA residue as the source of the PL shift.

In conjunction with PL measurements, reflectivity measurements are obtained from the same spot on the sample. Light from a white light source is focused though a 50× objective onto the sample and reflected intensity is measured. To minimize unwanted contributions from the substrate, the difference between the intensity measured from the bare substrate ($I_{off}$), and from the WS$_2$ sample ($I_{on}$) is obtained then normalized to the substrate intensity, $I^* = (I_{off}-I_{on})/I_{off}$. The peak reflectivity measured at 1.961 eV (2.015 eV) for as-WS$_2$ (PMMA x-WS$_2$) is nearly coincident with PL measurements, although there is a small difference in peak energy. This measured Stokes shift provides information regarding the doping level of TMDs, as Stokes shifts are known to increase monotonically with the doping level.[32] The small, ≤ 5meV Stokes shift observed in both the as-grown and transferred samples is comparable to or less than that observed in exfoliated WS$_2$ monolayers,[1,8] and indicates that a low doping level is achieved in as-grown material and maintained throughout the transfer process.

Raman spectra also show considerable change after PMMA transfer (Fig. 3 (c-h)). Measurements are performed with excitation energies of both 532 nm and 488 nm in an effort to reduce the resonant 2LA contribution. For both laser excitations, the relative intensities of the dominant $E^1_{2g}$, $A_{1g}$, and 2LA modes show considerable differences before and after transfer. We surmise that the transfer process increases the WS$_2$-to-substrate distance, leading to differences in optical interference, which in turn affects the Raman peak intensity.[33,34] These complicated interference effects are sensitive to both laser excitation energy as well as the energy of the Raman emission, and can thereby enhance one Raman peak while reducing the intensity of a second, as seen in Fig. 3(c,d). Therefore, it is instructive to carefully examine any alterations in peak positions (Fig. 3(e-h)).

The position of the out-of-plane A$_{1g}$ mode in TMD monolayers is sensitive to doping level, and red-shifting occurs with increasing electron concentration, while E$^1_{2g}$ is unaffected by doping levels.[3,35] Conversely only the E$^1_{2g}$ mode is impacted by strain, and exhibits a red-shift as a result of uniaxial and biaxial strain.[36,37] For both the 532 nm and 488 nm laser excitations there is a clear ~1 cm$^{-1}$ shift in the E$^1_{2g}$ peak (Fig. 3(e-f)), whereas the A$_{1g}$ peak position is unchanged (Fig. 3(g-h)). The observed shift is consistent with a reduction of tensile strain.[38,39] The tensile strain arises due to the differences in the thermal expansion coefficient of the monolayer WS$_2$ and the supporting substrate. As the substrate and WS$_2$ cool to room temperature from growth temperature (e.g. 825° C), the two contract at different rates, imparting strain to the WS$_2$ monolayer. The transfer process subsequently removes the strain. The presence of tensile strain is expected to reduce the band gap of WS$_2$ at a rate of 0.2 eV per % strain.[26,38] We therefore estimate the strain present in as-WS$_2$ on Si/SiO$_2$ to be ~0.25% based on the ~50 meV shift in PL emission measured before and after transfer.

Sample uniformity is investigated across representative ~30 μm triangles. Detailed maps (Fig. 4) show the PL intensity, peak position, and FHWM for as-grown

WS$_2$ on Si/SiO$_2$ compared to WS$_2$ synthesized on Si/SiO$_2$ and subsequently transferred to fresh Si/SiO$_2$ using both PMMA and PDMS transfer methods. The PL emission from as-WS$_2$ (Fig. 4(b-d)) exhibits considerable variation across the sample. The perimeter exhibits high PL intensity, with low intensity regions in the center that extend radially outward toward the three corners, similar to previous reports.[17,28] Despite the variation in PL intensity, the peak position and FWHM (Fig. 4c, d) are highly uniform across the sample, with average values of 1.964eV and 36 meV, respectively.

The PMMA x-WS$_2$ (Fig. 4(f-h)) exhibits a similar spatial pattern in PL intensity, with low intensity at the center and high intensity along the edges. Overall, the PL emission undergoes a blueshift of ~50meV, exhibiting an average value of 2.012eV, while the average FWHM is 32meV. The similarity in PL intensity patterns for as-grown and transferred samples combined with the uniformity in peak position and FWHM exclude local variations in strain or doping as the cause of the non-uniform PL intensity,[40] but instead points to structural defects as the source of non-uniformity. The PL intensity of PDMS x-WS$_2$ (Fig. 4(j-l)) is comparable to the PMMA transfer, with qualitatively similar PL intensity variations across the triangle, and uniform emission energy at 2.013eV and average FWHM of 36meV. The edges exhibit a slightly higher FWHM relative to the interior, most likely due to the tears, which are evident in the optical images. While some sample-to-sample variation is observed in emission energy following transfer, the overall shift to higher emission energy is highly repeatable, typically resulting in emission above 2.0 eV.

An investigation of the relationship between emission spectra and laser excitation power is presented in Fig. 5. Recent reports have shown that exposure to a high intensity laser can enhance trionic[8,41] emission as well as induce biexciton[42,43] emission in $WS_2$. As is evident from the normalized PL spectra of as-$WS_2$ emission (Fig. 5(a)), there is a single peak for all incident powers spanning over four orders of magnitude, consistent with emission from the neutral exciton ($X^0$). The integrated PL intensity is obtained within the range from 1.65 eV to 2.2 eV at each laser power (Fig. 5(d)) and is well described by a linear relationship (black line). The linearity is indicative of emission from the neutral exciton with an absence of exciton-exciton recombination. Additionally, the PL peak position of the as-$WS_2$ is nearly constant, showing only a slight decrease at the highest laser powers investigated, (Fig. 5(e)) likely due to heating effects.

Transferred samples exhibit considerably different behavior under identical excitation conditions. At low power, both PMMA (Fig. 5(b)) and PDMS (Fig. 5(c)) x-$WS_2$ are in the neutral excitonic state, $X^0$. As laser power steadily increases, a low energy shoulder emerges then quickly develops into the dominant emission peak. The lower emission energy and larger FWHM identify the second emission peak as the charged exciton, or trion (T).[41] The integrated PL intensity of transferred samples begins to deviate from the linear relationship above 1μW laser power, exhibiting lower integrated area as laser power increases (Fig. 5(d)). The decrease in emission is consistent with a transfer from $X^0$ to T dominated emission, as emission from charged trions is known to be lower in intensity due to increased non- radiative recombination mechanisms such as Auger recombination or exciton-exciton annihilation.[3,8,44] The position of the neutral exciton is nearly stable throughout the range of laser powers, undergoing a small redshift (<5meV) only at the highest powers. Such behavior is similar to that observed in as-WS$_2$ and suggests the heat transferred to the substrate is comparable for both as-WS$_2$ and x-WS$_2$. The trion in x-WS$_2$, however, exhibits a clear redshift with increasing laser power (Fig. 5(e)), resulting in an increasing separation between X$^0$ and T. The separation between neutral and charged excitons is linearly dependent on the Fermi level and is described by,

$$E_{X^0} - E_T = E_{b,T} + E_F$$

where $E_{X^0}$ is the energy position of the neutral exciton, $E_T$ is the position of the trion, $E_{b,T}$ is the trion binding energy, and $E_F$ is the Fermi level.[32,42] The PMMA (PDMS) transferred sample exhibits a separation of 26meV (30meV) at the 1μW excitation, providing an upper bound for the trion binding energy, consistent with previous results on mechanically exfoliated WS$_2$.[42,44]

Tunability between X$^0$ and T emission has previously been demonstrated in mechanically exfoliated monolayer TMDs following high powered laser exposure or annealing procedures. The likely source of the tunability is the removal or adsorption of p-type oxygen containing adsorbates.[41,45–49] We note that all our samples are measured in an ambient environment and are therefore exposed to similar concentrations of adsorbates. The stark contrast we observe between as-grown and transferred WS$_2$ thus suggests very different adsorption/desorption mechanisms are present on the two types of samples. In studies of the related 2D material graphene, it has been shown that surface deformations impact sample-adsorbate interactions. Specifically, binding at local sites becomes more energetically favorable.[50] Microscopic profile fluctuations are common in mechanically exfoliated or transferred materials due to transfer-induced wrinkles or trapped chemicals.[51–53] This effect may be significantly reduced for as-grown materials on a rigid substrate, explaining the lack of tunability on these types of samples. Additionally, differences in sample-substrate interactions are also expected to impact adsorption behaviors. Some mechanisms include variations in strain,[54–56] out-of-plane relaxation at binding sites,[57,58] and accessibility to top and bottom surfaces.[56]

Synthesis of $WS_2$ on substrates with different thermal expansion coefficients ($\alpha$) provides a means to independently modify the strain in $WS_2$ via the growth cooldown and subsequently monitor the sensitivity to laser power. Using the same procedure as reported above, we synthesize monolayer $WS_2$ on fused silica and c-plane sapphire. The coefficients of thermal expansion for silica and sapphire are 0.55 and 9.7 x $10^{-6}$/K, respectively. The thermal expansion coefficient for $WS_2$ is expected to be in the range $7 < \alpha < 10 \times 10^{-6}$/K based on theoretical investigations [59] and experimentally measured bulk values.[60] As with the growth of $WS_2$ on $Si/SiO_2$, triangular growth is evident on both fused silica and sapphire substrates (Fig. 6). Maps of the PL peak positions are presented in Fig. 6(a-b). $WS_2$ synthesized on fused silica exhibits uniform emission energy across the triangle, with an average PL peak position of 1.94 eV. In contrast, $WS_2$ on sapphire results in the considerably higher average emission energy of 1.99 eV, and a large variation in peak position (1.95-2.02 eV). This large variation on sapphire could arise from factors such as substrate impurities or water intercalation[25] and is under investigation.

The energy position of PL emission and associated Raman spectra (supplementary information) demonstrate that synthesis on the low-α fused silica substrates induce considerable strain in $WS_2$ whereas growth on sapphire approaches a strain-free system. Power-dependent PL spectra are shown in Fig. 6 (c,d) for as-grown samples on both fused silica and sapphire. Based on a value of 2.105 eV for an unstrained $WS_2$ monolayer, the $WS_2$ on fused silica (sapphire) exhibits ~0.39% (0.08%) strain. In both cases, the spectral shape and position are unaffected by laser power. Additionally, the integrated area of the PL emission follows a linear behavior for all laser powers used, indicating the excitonic behavior is not modified with increasing laser power. Therefore, while as-grown $WS_2$ on $Si/SiO_2$, sapphire, and fused silica exhibit distinctly different strain amounts, optical control of the excitonic state is absent, suggesting strain has minimal impact on adsorption and desorption mechanisms of $O_2$ containing species.

**Conclusion**

In conclusion, we have demonstrated that the choice of growth substrate determines the PL emission energy of the neutral exciton, $X^0$. The emission energy is governed by strain in the as-grown $WS_2$ and arises from differences in thermal expansion coefficient between $WS_2$ and the supporting substrate. The strain can be removed by transferring the sample off of the growth substrate. This process results in the emission shifting to higher energy. We also observe a different response to laser power for as-grown versus transferred samples. Transferred samples exhibit extreme sensitivity to laser power, with $X^0$ emission dominating at low laser power, but transitioning to T emission as power is increased. As-grown materials show no variation in spectral shape or position of PL as laser power is increased by more than four orders of magnitude. We speculate that the distinctly different power-dependent behaviors result primarily from different adsorption and/or desorption mechanisms in as-grown and transferred $WS_2$. Differences in interfacial effects such as electron transfer between $WS_2$ and supporting substrates could also influence the observed behaviors. The more intimate $WS_2$-substrate contact present in as-$WS_2$ may promote more efficient electron transfer from $WS_2$ to the substrate, hindering trion formation. Theoretical modeling and continued experimental investigation are necessary to identify the precise source of disparity observed in as-$WS_2$ and x-$WS_2$.

The moderate laser powers used in this work are comparable to conditions typically utilized for optical characterization of TMDs. This highlights that care is required to prevent laser excitation-induced effects in transferred materials. Additionally, we speculate that investigations involving chemical sensing and defect passivation will be distinctly different for as-grown and transferred/exfoliated materials, as both are intimately related to adsorption and desorption mechanisms. Finally, we note this work is the first to report such dramatically different power-dependent behavior for as-grown and transferred $WS_2$. This study should therefore stimulate further experimental and theoretical investigations in the field.

## Methods
**Sample preparation "PMMA transfer".** A thin layer of polymethyl methacrylate (PMMA) is spun onto the surface of the entire growth substrate then submerged in buffered oxide etchant. After several hours, the oxide layer is removed, freeing the WS$_2$/PMMA film from the growth substrate. The sample is subsequently transferred to H$_2$O to rinse chemical etchants, where the fresh Si/SiO$_2$ is used to lift the film out of the water. A 2000 rpm spin and 150°C bake improve the uniformity and adhesion to the substrate, after which the PMMA is dissolved in acetone.

**Sample preparation "PDMS transfer".** A polydimethylsiloxane (PDMS) stamp coated with polycarbonate (PC) is carefully brought into contact with the desired monolayer region. The van der Waals forces between the PC are strong enough to remove the WS$_2$ from the supporting substrate. Once on the stamp, the WS$_2$ can be positioned or rotated to a desired angle before being brought into contact and released onto the desired substrate with a gentle thermal anneal.

**Acknowledgements**
Core programs at NRL and the NRL Nanoscience Institute supported this work. This work was supported in part by the Air Force Office of Scientific Research under contract number AOARD 14IOA018-134141. The work at Ohio State was supported by NSF (DMR-1310661). The authors acknowledge use of facilities in the NRL Nanoscience Institute and thank Dean St. Amand and Walter Spratt for technical support.

**Author Contributions**
K.M. performed the CVD synthesis of samples. K.M. and A.H. performed optical characterization. S.S. performed PDMS stamp transfer.  K.M., A.T., S.S., R.K., and B.J discussed the results and contributed to the manuscript.

**Additional Information**
**Supplementary Information**: Additional details are available at
http://www.nature.com/srep

**Competing financial interests:** The authors declare no competing financial interests.

**Figure captions**

**Figure 1: Characterization of monolayer WS$_2$ synthesized on Si/SiO$_2$.** (a) Optical microscope images display randomly oriented equilateral triangular growth. (b) The AFM image is obtained on a representative WS$_2$ triangle. A height profile (inset) is measured along the dashed line and shows a step height of ~0.8 nm. (c) Raman spectroscopy and (d) photoluminescence measurements confirm monolayer WS$_2$.

**Figure 2: Schematic of the two transfer methods.** (a) For PMMA transfer, the sample is coated with a thin layer of PMMA then submerged in BOE to remove the SiO$_2$. Once fully etched, the film is rinsed in H$_2$O then picked up with the target substrate. An acetone soak removes the PMMA. An optical image of PMMA transferred WS$_2$ exhibits a uniform, clean, triangular shape. (b) For the PDMS transfer, a PDMS/PC film is carefully brought into contact with the desired WS$_2$ then retracted. This moves the WS$_2$ from Si/SiO$_2$ onto the PDMS/PC film. The PDMS/PC/WS$_2$ stack is then placed onto clean Si/SiO$_2$. The PDMS stamp is retracted, leaving the PC film on the top surface of WS$_2$, which is then dissolved in chloroform. An optical image following PDMS transfer is shown on the right.

**Figure 3: PL, reflectivity, and Raman characterization before and after PMMA transfer.** (a) PL emission and (b) reflectivity shift to higher energy following transfer. A ≤5 meV difference in PL and reflectivity spectra is measured for both as-grown and transferred WS$_2$. The vertical red and blue dashed lines indicate the peak PL position and highlight the small Stokes shift. Raman spectra are measured using (c) 532 nm and (d) 488 nm excitation. As-grown and transferred spectra are offset for clarity. The relative intensities of $E^1_{2g}$, $A_{1g}$, and 2LA vary before and after. The intensity of the Si substrate peak (520.7 cm$^{-1}$) is unaffected. (e,f) The transfer process shifts the in-plane Raman mode to higher wavenumber while out-of-plane (g,h) remains constant. Schematic diagrams of the $E^1_{2g}$ and $A_{1g}$ Raman modes are displayed in the inset of (f) and (h), respectively. The changes in optical properties indicate the transfer process removes tensile strain

**Figure 4: Optical images and PL maps comparing as-grown and transferred WS$_2$.** (a) Optical microscopy of as-grown WS$_2$ and corresponding maps of (b) PL intensity, (c) PL emission position, and (d) FWHM. (e) Optical microscopy of PMMA transferred WS$_2$ and corresponding maps of (f) PL intensity, (g) PL emission position, and (h) FWHM. (i) Optical microscopy of PDMS transferred WS$_2$ and corresponding maps of (j) PL intensity, (k) PL emission position, and (l) FWHM. Similar patterns in PL intensity are observed for all three samples, with low intensity at the center of the triangle and directed outward toward the three corners. PMMA and PDMS transferred samples have noticeably higher emission energy, resulting from the removal of strain.

**Figure 5: Laser power dependence in ambient conditions.** (a) For as-grown $WS_2$, the spectral shape of PL is unaffected by increased laser power, with emission occurring from the neutral exciton, $X^0$, for all excitation conditions. In (b) PDMS and (c) PMMA transferred samples, $X^0$ emission dominates at low powers, but the trion, T, dominates at high power excitation. (d) PL integrated area from 1.65 to 2.2 eV and (e) peak positions for the three samples.

**Figure 6: Characterization of monolayer $WS_2$ synthesized on various substrates.** (a,b) Maps of PL position for $WS_2$ monolayers grown on fused silica and sapphire, respectively. The average peak position on silica (sapphire) is 1.94 ev (1.99 eV). The dissimilar thermal expansion coefficients of silica and sapphire lead to different strain in $WS_2$ and the differences in peak position. The position and spectral shape of PL emission is unaffected by increased laser power for both (c) silica and (d) sapphire. (e,f) Integrated area is linearly related to laser power for both silica and sapphire. The insets of (c) and (d) show optical images of the mapped $WS_2$ monolayers. The scale bar is 10 μm.

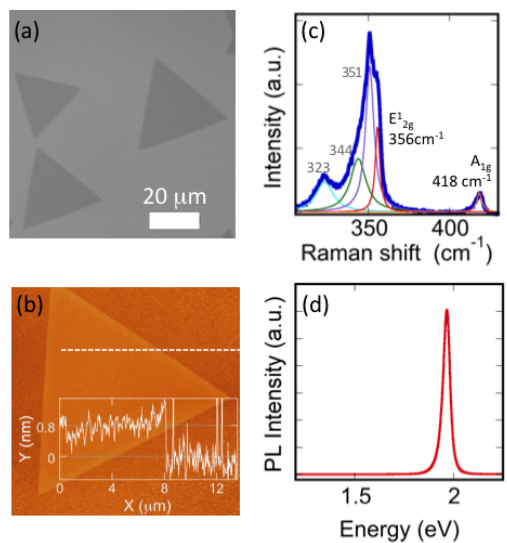

**Figure 1: Characterization of monolayer WS$_2$ synthesized on Si/SiO$_2$.** (a) Optical microscope images display randomly oriented equilateral triangular growth. (b) The AFM image is obtained on a representative WS$_2$ triangle. A height profile (inset) is measured along the dashed line and shows a step height of ~0.8 nm. (c) Raman spectroscopy and (d) photoluminescence measurements confirm monolayer WS$_2$.

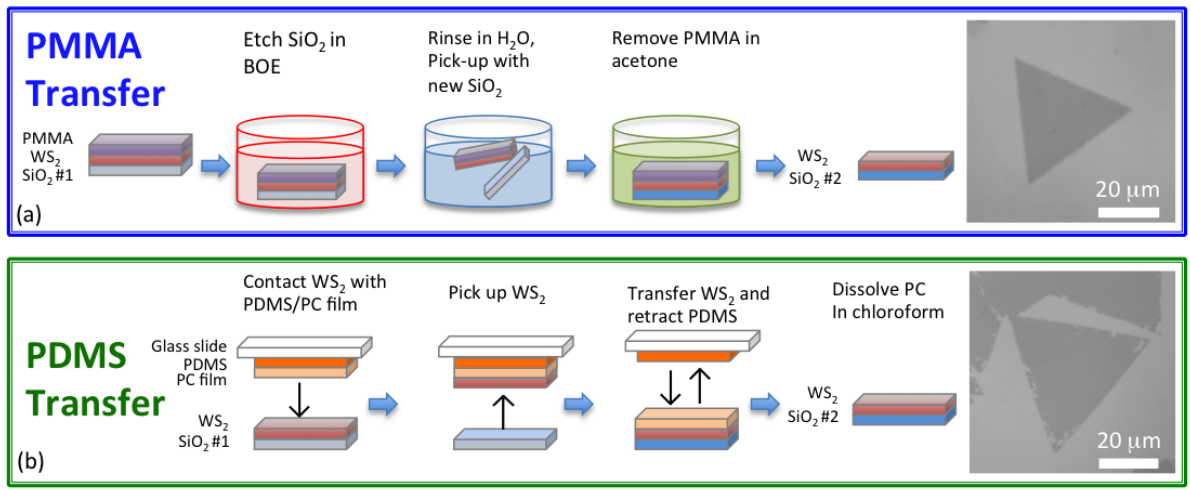

**Figure 2: Schematic of the two transfer methods.** (a) For PMMA transfer, the sample is coated with a thin layer of PMMA then submerged in BOE to remove the $SiO_2$. Once fully etched, the film is rinsed in $H_2O$ then picked up with the target substrate. An acetone soak removes the PMMA. An optical image of PMMA transferred $WS_2$ exhibits a uniform, clean, triangular shape. (b) For the PDMS transfer, a PDMS/PC film is carefully brought into contact with the desired $WS_2$ then retracted. This moves the $WS_2$ from $Si/SiO_2$ onto the PDMS/PC film. The PDMS/PC/$WS_2$ stack is then placed onto clean $Si/SiO_2$. The PDMS stamp is retracted, leaving the PC film on the top surface of $WS_2$, which is then dissolved in chloroform. An optical image following PDMS transfer is shown on the right.

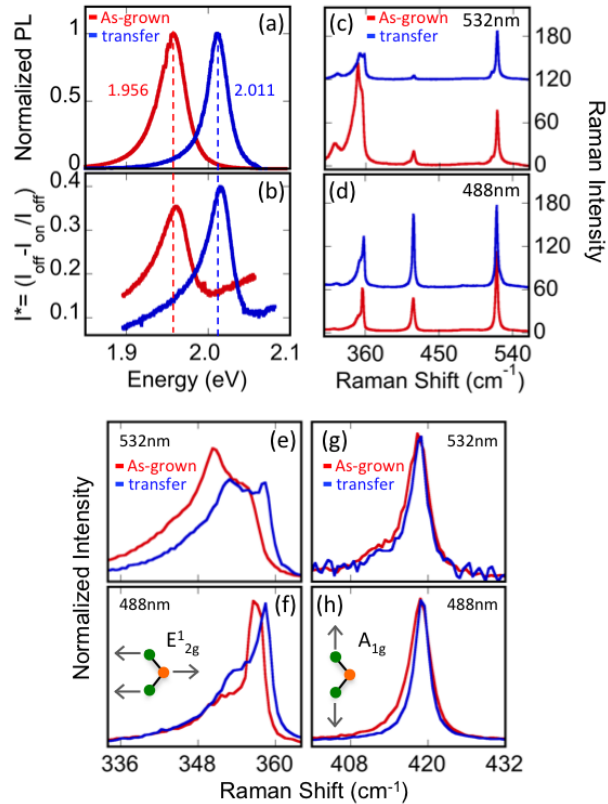

**Figure 3: PL, reflectivity, and Raman characterization before and after PMMA transfer.** (a) PL emission and (b) reflectivity shift to higher energy following transfer. A ≤5 meV difference in PL and reflectivity spectra is measured for both as-grown and transferred $WS_2$. The vertical red and blue dashed lines indicate the peak PL position and highlight the small Stokes shift. Raman spectra are measured using (c) 532 nm and (d) 488 nm excitation. As-grown and transferred spectra are offset for clarity. The relative intensities of $E^1_{2g}$, $A_{1g}$, and 2LA vary before and after. The intensity of the Si substrate peak (520.7 $cm^{-1}$) is unaffected. (e,f) The transfer process shifts the in-plane Raman mode to higher wavenumber while out-of-plane (g,h) remains constant. Schematic diagrams of the $E^1_{2g}$ and $A_{1g}$ Raman modes are displayed in the inset of (f) and (h), respectively. The changes in optical properties indicate the transfer process removes tensile strain

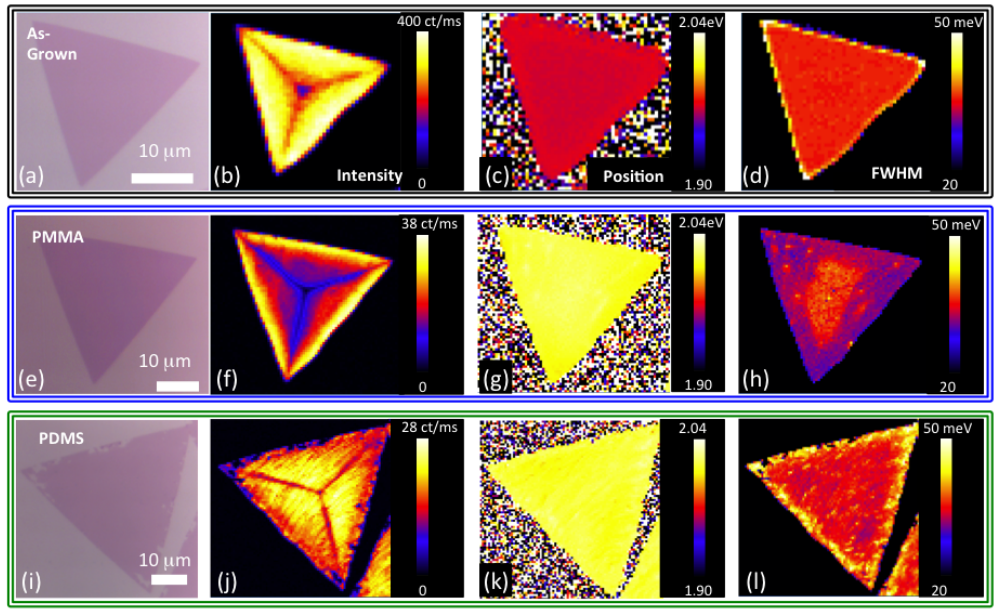

**Figure 4: Optical images and PL maps comparing as-grown and transferred WS$_2$.** (a) Optical microscopy of as-grown WS$_2$ and corresponding maps of (b) PL intensity, (c) PL emission position, and (d) FWHM. (e) Optical microscopy of PMMA transferred WS$_2$ and corresponding maps of (f) PL intensity, (g) PL emission position, and (h) FWHM. (i) Optical microscopy of PDMS transferred WS$_2$ and corresponding maps of (j) PL intensity, (k) PL emission position, and (l) FWHM. Similar patterns in PL intensity are observed for all three samples, with low intensity at the center of the triangle and directed outward toward the three corners. PMMA and PDMS transferred samples have noticeably higher emission energy, resulting from the removal of strain.

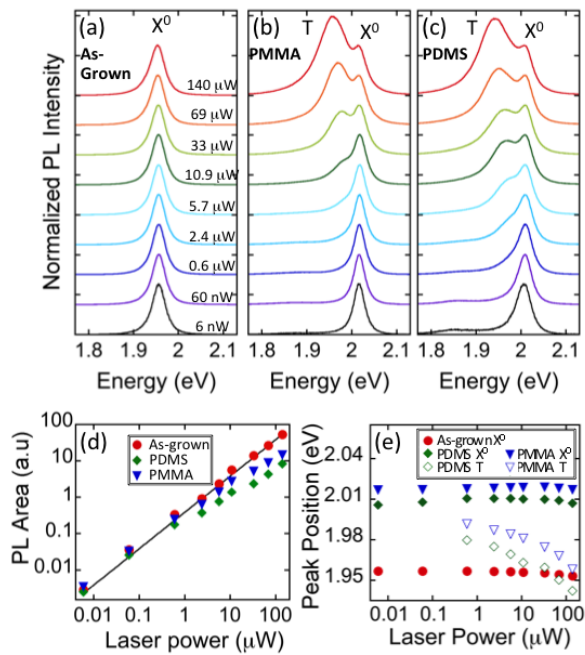

**Figure 5: Laser power dependence in ambient conditions**. (a) For as-grown $WS_2$, the spectral shape of PL is unaffected by increased laser power, with emission occurring from the neutral exciton, $X^0$, for all excitation conditions. In (b) PDMS and (c) PMMA transferred samples, $X^0$ emission dominates at low powers, but the trion, T, dominates at high power excitation. (d) PL integrated area from 1.65 to 2.2 eV and (e) peak positions for the three samples.

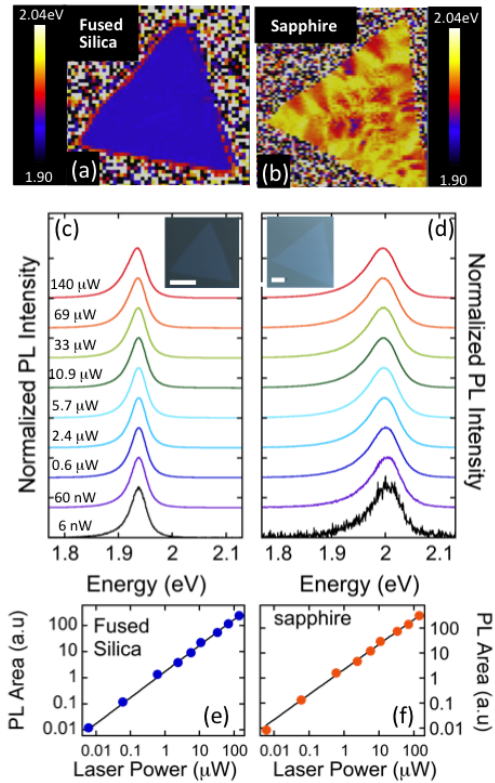

**Figure 6: Characterization of monolayer WS$_2$ synthesized on various substrates.** (a,b) Maps of PL position for WS$_2$ monolayers grown on fused silica and sapphire, respectively. The average peak position on silica (sapphire) is 1.94 ev (1.99 eV). The dissimilar thermal expansion coefficients of silica and sapphire lead to different strain in WS$_2$ and the differences in peak position. The position and spectral shape of PL emission is unaffected by increased laser power for both (c) silica and (d) sapphire. (e,f) Integrated area is linearly related to laser power for both silica and sapphire. The insets of (c) and (d) show optical images of the mapped WS$_2$ monolayers. The scale bar is 10 μm.